\documentclass[aps,prb,superscriptaddress,reprint]{revtex4-1}

\usepackage{amsmath}
\usepackage{amssymb}
\usepackage{graphicx}
\usepackage{epstopdf}
\usepackage{dcolumn}
\usepackage{bm}
\usepackage{cases}
\usepackage{mathtools}
\usepackage{dcolumn}
\usepackage{color}
\usepackage{booktabs}
\usepackage{placeins}

\DeclarePairedDelimiter\bra{\langle}{\rvert}
\DeclarePairedDelimiter\ket{\lvert}{\rangle}

\renewcommand\vec[1]{{\bf #1}}

\begin{document}
 
\title{All-electrical manipulation of silicon spin qubits with tunable spin-valley mixing}
 
\author{L\'eo Bourdet}
\author{Yann-Michel Niquet}
\email{yniquet@cea.fr}
\affiliation{University Grenoble Alpes, CEA, INAC-MEM, 38000 Grenoble, France}

\begin{abstract}
We show that the mixing between spin and valley degrees of freedom in a silicon quantum bit (qubit) can be controlled by a static electric field acting on the valley splitting $\Delta$. Thanks to spin-orbit coupling, the qubit can be continuously switched between a spin mode (where the quantum information is encoded into the spin) and a valley mode (where the the quantum information is encoded into the valley). In the spin mode, the qubit is more robust with respect to inelastic relaxation and decoherence, but is hardly addressable electrically. It can however be brought into the valley mode then back to the spin mode for electrical manipulation. This opens new perspectives for the development of robust and scalable, electrically addressable spin qubits on silicon. We illustrate this with tight-binding simulations on a so-called ``corner dot'' in a silicon-on-insulator device where the confinement and valley splitting can be independently tailored by a front and a back gate.
\end{abstract}

\maketitle


Silicon\cite{Zwanenburg13} is an attractive material for solid-state quantum bits (qubits) owing to its mature technology and very long spin coherence times.\cite{Tyryshkin12} As a matter of fact, high fidelity single qubits and two qubit gates have already been demonstrated in silicon.\cite{Veldhorst15b,Takeda16,Veldhorst14}

The spin of electrons and holes in silicon quantum dots (QDs) is routinely manipulated with radio-frequency (RF) magnetic fields (Electron Spin Resonance).\cite{Veldhorst14,Veldhorst15,Laucht15} RF magnetic fields can, however, hardly be applied locally. 
In the prospect of controlling a large number of qubits, it may be less demanding to manipulate spins with the RF electric field from a local gate (Electric Dipole Spin Resonance or EDSR). This calls for a mechanism that couples the orbital motion of the electron with its spin. One possible strategy is to introduce micro-magnets that create a gradient of magnetic field in the QD, giving rise to an effective spin-orbit interaction.\cite{Pioro-Ladriere08,Kawakami14} However, in order to achieve compact and simple designs, it is more attractive to rely on the ``intrinsic`` spin-orbit coupling (SOC) of the host material. SOC-mediated EDSR has first been demonstrated for electrons and holes in III-V QDs,\cite{Nowack07,Nadj-Perge10,Pribiag13} then for holes in silicon QDs.\cite{Maurand16} It is much more challenging for electrons in silicon QDs, because SOC is very weak in the conduction band of Si.\cite{Huang17} Yet SOC-mediated EDSR has been achieved very recently in the ``corner dots'' of silicon-on-insulator (SOI) nanowire channels.\cite{Corna17}

The underlying mechanism relies on the extraordinary rich and complex physics of electrons in silicon.\cite{Zwanenburg13,Nestoklon06} Bulk silicon is an indirect bandgap material with six degenerate conduction band valleys. This degeneracy is completely lifted in silicon QDs. Structural and electric confinement indeed leaves only two low-lying valleys $v_1$ and $v_2$ separated by a valley splitting energy $\Delta$,\cite{Sham79,Saraiva09,Friesen10,Culcer10} which ranges from a few $\mu$eVs to a few meVs.\cite{Goswami07,Yang13,Corna17,Mi17} At a critical magnetic field $B_{\rm A}$, the spin down state of valley $v_2$ crosses the spin up state of valley $v_1$, and get mixed by the weak SOC.\cite{Hao14,Scarlino17} This allows for electrically driven transitions between the $\ket{v_1,\downarrow}$ state and the mixed $\ket{v_1,\uparrow}$/$\ket{v_2,\downarrow}$ state, thanks to the existence of a non-zero dipole matrix element between $\ket{v_1,\downarrow}$ and $\ket{v_2,\downarrow}$.\cite{Corna17} However, the spin relaxation time $T_1$ and spin coherence time $T_2$ are expected to be shorter near that anti-crossing due to the enhanced coupling of the spin to electric noise and phonons.\cite{Yang13,Huang14} 

The valley splitting $\Delta$ can be controlled over a wide range by external electric fields.\cite{Goswami07,Yang13} This is particularly the case in SOI devices, which feature an additional substrate back gate, but also holds in carefully designed multi-gate planar structures. In this letter, we show with tight-binding simulations how multiple gates can be efficiently used to tune the silicon QD and sweep it across the anti-crossing point. The qubit can then be adiabatically switched between one ``valley'' mode\cite{Schoenfield17} that can be manipulated with RF electric fields, and one ``spin'' mode\cite{Loss98} whose evolution is much less sensitive to electric noise and phonons. Such a scheme allows for the implementation of robust and electrically addressable silicon spin qubits.\cite{Kloeffel13} We first review the theory of SOC-mediated EDSR, then discuss the control of the valley splitting, and finally present the spin manipulation protocol.


\begin{figure}
\includegraphics[width=.90\columnwidth]{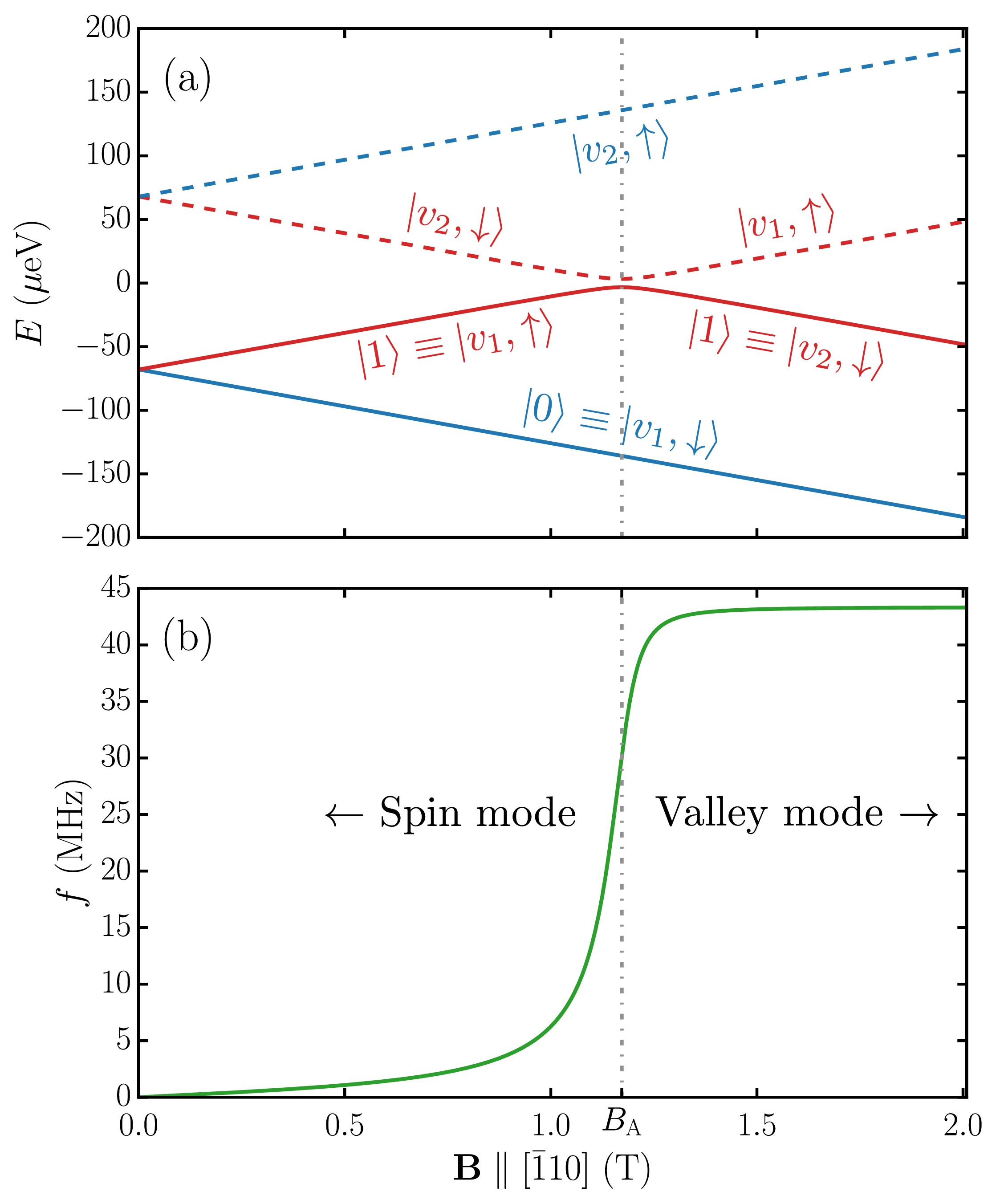} 
\caption{(a) Energy levels of a silicon QD in a magnetic field $B$. The solid blue line is the energy of the $\ket{v_1,\downarrow}$ state, the dotted blue line the energy of the $\ket{v_2,\uparrow}$ state, and the solid and dashed red lines the energies of the $\ket{\psi_-}$ and $\ket{\psi_+}$ states (which are mixtures of the $\ket{v_1,\uparrow}$ and $\ket{v_2,\downarrow}$ states that anti-cross at $B=B_{\rm A}=1.172$ T). (b) Computed Rabi frequency for the transition between $\ket{0}\equiv\ket{v_1,\downarrow}$ and $\ket{1}\equiv\ket{\psi_-}$. The parameters of the model are $\Delta=136$ $\mu$eV, $|C_{v_1v_2}|=3.25$ $\mu$eV, and $|D_{v_1v_2}|=179.26$ $\mu$V/V. They have been extracted from tight-binding simulations on the device of Fig. \ref{figDEV} at $V_{\rm fg}=0.1$ V and $V_{\rm bg}=0$ V. The amplitude of the RF excitation on the front gate is $\delta V_{\rm fg}=1$ mV.}
\label{figSV}
\end{figure}

The theory of spin-orbit mediated EDSR in the conduction band of silicon has been discussed in Ref. \onlinecite{Corna17}. We recall the main elements here.\cite{Yang13,Huang14}

We consider a silicon QD strongly confined along the $z$ direction so that the low-energy levels belong to the $\Delta_{\pm z}$ valleys. In the absence of valley coupling, the ground-state level is fourfold degenerate (twice for spins and twice for valleys). Valley coupling\cite{Zwanenburg13,Sham79,Saraiva09,Friesen10,Culcer10} splits this level into two spin-degenerate states $\ket{v_1,\sigma}$ and $\ket{v_2,\sigma}$ with energies $E_1$ and $E_2$, separated by the valley splitting energy $\Delta=E_2-E_1$ ($\sigma=\,\uparrow,\downarrow$ is the spin index). The remaining spin degeneracy can be lifted by an external magnetic field $\vec{B}$. The energy of state $\ket{v_n,\sigma}$ is $E_{n,\sigma}=E_n+\frac{1}{2}g\mu_B B\sigma$, where $g\simeq 2$ is the gyro-magnetic factor of the electrons (the spin being quantified along $\vec{B}$). 

The energy $E_{n,\sigma}$ of the spin-valley states is plotted as a function of $B$ in Fig. \ref{figSV}a. The states $\ket{v_1,\uparrow}$ and $\ket{v_2,\downarrow}$ are mixed by SOC and anti-cross at magnetic field $B=B_{\rm A}=\Delta/(g\mu_B)$. The energy of the upper (dashed red) and lower (solid red) branch of the anti-crossing read:
\begin{equation}
E_\pm=\frac{1}{2}(E_1+E_2)\pm\frac{1}{2}\sqrt{(\Delta-g\mu_B B)^2+4|C_{v_1v_2}|^2}\,,
\end{equation}
where $C_{v_1v_2}=\bra{v_2,\uparrow}H_{\rm SOC}\ket{v_1,\downarrow}=-\bra{v_1,\uparrow}H_{\rm SOC}\ket{v_2,\downarrow}$ is the matrix element of the spin-orbit coupling Hamiltonian $H_{\rm SOC}$ between valleys $v_1$ and $v_2$. The eigenstates of the upper and lower branch are, respectively:
\begin{subequations}
\begin{align}
\ket{\psi_+}&=\alpha\ket{v_1,\uparrow}+\beta\ket{v_2,\downarrow} \\
\ket{\psi_-}&=\beta\ket{v_1,\uparrow}-\alpha^*\ket{v_2,\downarrow}
\end{align}
\end{subequations}
with:
\begin{subequations}
\begin{align}
\alpha(\varepsilon)&=\frac{2C_{v_1v_2}^*}{\sqrt{\varepsilon^2+4|C_{v_1v_2}|^2}} \\
\beta(\varepsilon)&=\frac{\varepsilon}{\sqrt{\varepsilon^2+4|C_{v_1v_2}|^2}}
\end{align}
\end{subequations}
and:
\begin{equation}
\varepsilon=\Delta-g\mu_B B+\sqrt{(\Delta-g\mu_B B)^2+4|C_{v_1v_2}|^2}\,.
\end{equation}
Note that $|\alpha|=|\beta|=1/\sqrt{2}$ at $B=B_{\rm A}$, which highlights the strong mixing between spin and valley degrees of freedom near the anti-crossing. Although the states $\ket{v_1,\downarrow}$ and $\ket{v_2,\uparrow}$ do not anti-cross, we must for consistency account for a very small mixing by SOC (otherwise the Rabi frequency would not vanish\cite{Corna17} when $B\to0$) and introduce:
\begin{subequations}
\begin{align}
\ket{\psi_+^\prime}&=\alpha^\prime\ket{v_1,\downarrow}+\beta^\prime\ket{v_2,\uparrow} \\
\ket{\psi_-^\prime}&=\beta^\prime\ket{v_1,\downarrow}-\alpha^{\prime*}\ket{v_2,\uparrow}
\end{align}
\end{subequations}
where $\alpha^\prime\equiv-\alpha^*(\varepsilon^\prime)$, $\beta^\prime\equiv\beta(\varepsilon^\prime)$, and $\varepsilon^\prime=\Delta+g\mu_B B+\sqrt{(\Delta+g\mu_B B)^2+4|C_{v_1v_2}|^2}$ ($\alpha^\prime\simeq 0$, $\beta^\prime\simeq 1$ whatever $B$).

\begin{figure}
\includegraphics[width=.66\columnwidth]{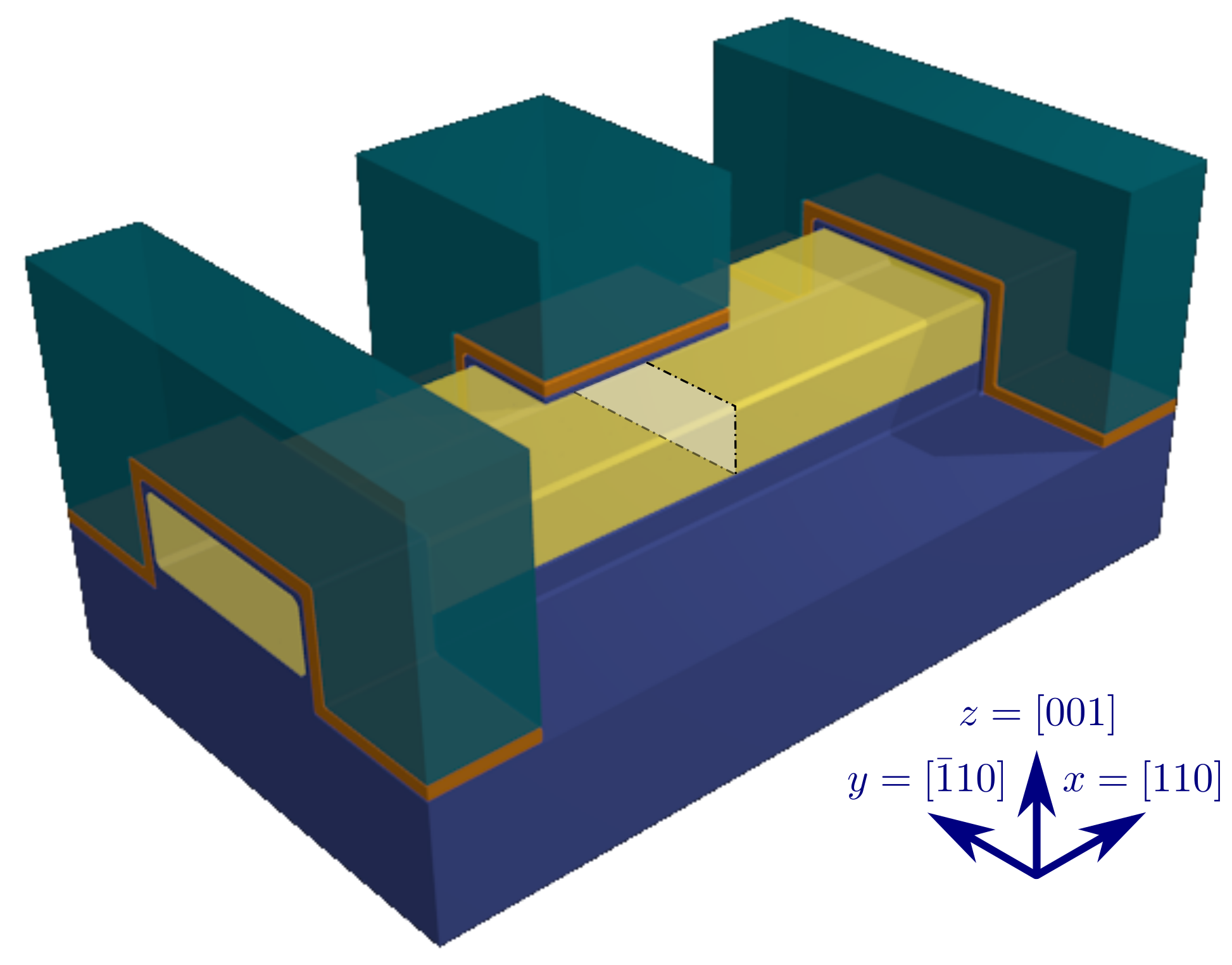} 
\caption{Schematics of the device. The $[110]$-oriented, $10\text{\ nm}\times 30\text{\ nm}$ silicon channel is colored in yellow; it lies on a 25 nm thick buried oxide (dark blue) with a doped silicon back gate beneath. The 30 nm long front gate (light blue) overlaps half of the channel; the front gate stack is made of 1 nm of SiO$_2$ and 2 nm of HfO$_2$ (brown). The two other lateral gates (also light blue) mimic adjacent qubits. They are biased at $V=0$ V throughout the letter. The gray area enclosed by the dashed lines is the cross section for wave function plots in Fig. \ref{figDELTA}}
\label{figDEV}
\end{figure}

\begin{figure*}
\includegraphics[width=.95\textwidth]{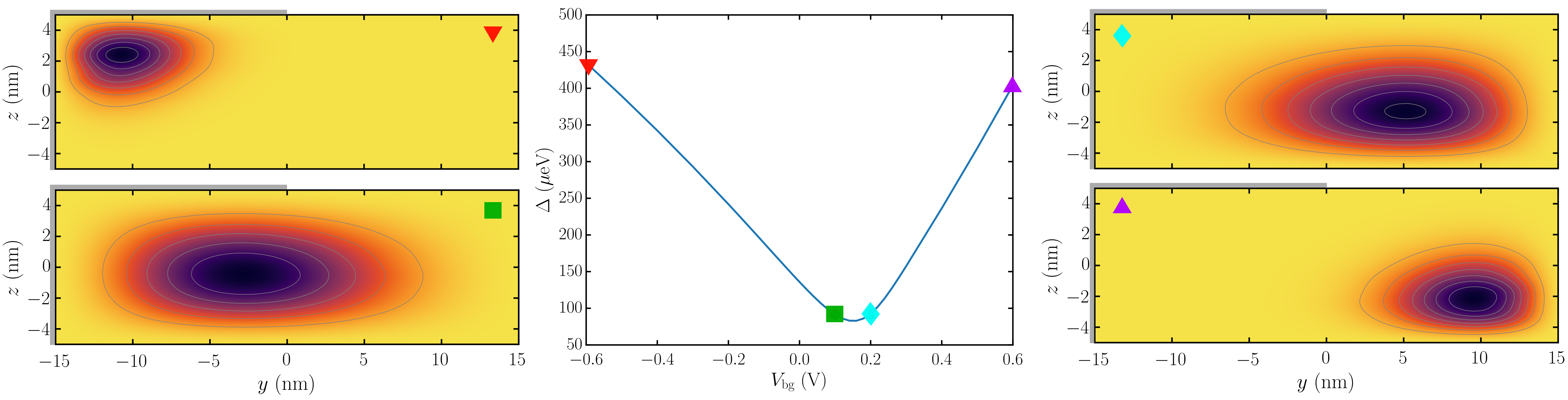} 
\caption{Valley splitting $\Delta$ as a function of the back gate voltage $V_{\rm bg}$ ($V_{\rm fg}=0.1$ V). The valley splitting shows a minimum $\Delta^{\rm min}=83$ $\mu$eV at $V_{\rm bg}^{\rm min}=0.15$ V, separating two domains where $\Delta$ depends almost linearly on $V_{\rm bg}$. The squared wave function of the ground state $\ket{0}$ is plotted in the $(yz)$ cross section grayed in Fig. \ref{figDEV} at the four bias points labeled by the markers ($\ket{1}$ shows an almost equivalent localization). The thick gray lines outline the position of the front gate. For $V_{\rm bg}\ll V_{\rm bg}^{\rm min}$, the electron is trapped near the top interface; while for $V_{\rm bg}\gg V_{\rm bg}^{\rm min}$ the electron is trapped near the BOX interface. For $V_{\rm bg}\simeq V_{\rm bg}^{\rm min}$, the electron sits in between the two interfaces.}
\label{figDELTA}
\end{figure*}

We are specifically interested in making a qubit based on states $\ket{0}=\ket{\psi_-^\prime}\simeq\ket{v_1,\downarrow}$ and $\ket{1}=\ket{\psi_-}$. Qubit rotations are driven by a RF modulation on a front gate voltage $V_{\rm fg}$. The Rabi frequency for the resonant transition between states $\ket{0}$ and $\ket{1}$ is then:
\begin{equation}
hf=e\delta V_{\rm fg}\left|\bra{\psi_-^\prime}D\ket{\psi_-}\right|=e\delta V_{\rm fg}|\alpha^*\beta^\prime+\alpha^\prime\beta||D_{v_1v_2}| 
\label{eqrabid}
\end{equation}
where $\delta V_{\rm fg}$ is the amplitude of the RF signal ($\delta V_{\rm fg}=1$ meV hereafter), $D(\vec{r})=\partial V_t(\vec{r})/\partial V_{\rm fg}$ is the derivative of the total potential $V_t(\vec{r})$ in the device with respect to $V_{\rm fg}$, and $D_{v_1v_2}=\bra{v_1,\sigma}D\ket{v_2,\sigma}$ is the gate coupling matrix element between valleys $v_1$ and $v_2$.

$f$ is plotted as a function of magnetic field in Fig. \ref{figSV}b, for values of $\Delta$, $D_{v_1v_2}$ and $C_{v_1v_2}$ extracted from tight-binding simulations on the device of Fig. \ref{figDEV} (see later discussion). For $B\ll B_{\rm A}$, $\ket{1}\sim\ket{v_1,\uparrow}$, so that the device is an almost ``pure spin'' qubit,\cite{Loss98} which is hardly addressable electrically. When increasing $B$, $\ket{1}$ admixes a growing fraction of $\ket{v_2,\downarrow}$, which is coupled to the ground-state $\ket{0}=\ket{v_1,\downarrow}$ by the RF electric field, allowing for Rabi oscillations (mixed spin/valley qubit). For $B\gg B_{\rm A}$, $\ket{1}\sim\ket{v_2,\downarrow}$, so that the device eventually becomes an almost ``pure valley'' qubit.\cite{Schoenfield17} The maximum Rabi frequency in this regime, $f_{\rm max}=e\delta V_{\rm fg}|D_{v_1v_2}|/h$, is therefore limited by the gate coupling matrix element $D_{v_1v_2}$. The width of the transition near $B=B_{\rm A}$ is controlled by the SOC matrix element $C_{v_1v_2}$, which sets the anti-crossing gap $E_{\rm SOC}=2|C_{v_1v_2}|$ at $B=B_{\rm A}$. The Rabi frequency may also depend on the orientation of the magnetic field (as the spin is quantized along $\vec{B}$ in the definition of $C_{v_1v_2}$).


The signatures of this spin resonance mechanism have been observed in a silicon nanowire device.\cite{Corna17} A model for this device is shown in Fig. \ref{figDEV}. The quantum dot is defined by a central gate on a silicon channel with rectangular cross section etched on a SOI substrate. The gate overlaps only half of the channel. The electrons are hence confined in ``corner dots'' at the edge of the channel covered by the gate.\cite{Voisin14} As discussed in Ref. \onlinecite{Corna17}, the formation of such low-symmetry dots is a key ingredient of the present spin resonance mechanism. Indeed, $C_{v_1v_2}$ is zero when the magnetic field $\vec{B}$ is perpendicular to a mirror plane; as an illustration, the Rabi frequency measured in Ref. \onlinecite{Corna17} is minimal when $\vec{B}$ is along the nanowire, and maximal when $\vec{B}$ is perpendicular to the nanowire, because $(yz)$ is a mirror plane in Fig. \ref{figDEV}. Consequently, SOC is inefficient in highly symmetric dots with more than one symmetry plane.

As discussed above, the Rabi frequency is maximal beyond the anti-crossing between $\ket{v_1,\uparrow}$ and $\ket{v_2,\downarrow}$ and can reach a few tens to a hundred of MHz depending on the device design and disorder.\cite{Corna17} This is much larger than the Rabi frequencies achieved with extrinsic elements such as micro-magnets. However, a QD operating in this regime would not make a good qubit. Indeed, the vicinity of the anti-crossing and the valley mode beyond are known to be ``hot spots'' for relaxation\cite{Yang13,Huang14} (shorter $T_1$) and decoherence (shorter $T_2$ due to enhanced sensitivity to charge and gate noise). Also, the strong mixing between $\ket{v_1}$ and $\ket{v_2}$ states near the anti-crossing may complicate the management of exchange interactions between neighboring qubits.

It would, therefore, be highly desirable to bring the qubit in the valley regime (beyond the anti-crossing) in order to manipulate its state electrically, then back to the spin regime (before the anti-crossing) once rotations are completed. The transitions between the two regimes must be performed adiabatically in order to achieve well defined operations.

The most obvious way to tune the spin/valley mixing is to vary the amplitude of the external magnetic field $\vec{B}$ (see Fig. \ref{figSV}). However, fast variations of $B$ are unrealistic, and would affect all qubits at once. An other way is to control the valley splitting with the gate(s). It has already been demonstrated that the valley splitting at a Si/SiO$_2$ interface depends on the electric field at that interface,\cite{Goswami07,Yang13} and can span orders of magnitudes. Nonetheless, it is generally difficult to control both the confinement potential and the vertical electric field with a set of front gates, which limits the range of achievable valley splittings. In SOI devices, the presence of both a front and a back gate allows, in principle, to decouple the confinement potential from the vertical electric field, and to implement electrical manipulation schemes based on the control of the valley splitting more easily.


\begin{figure}
\includegraphics[width=.90\columnwidth]{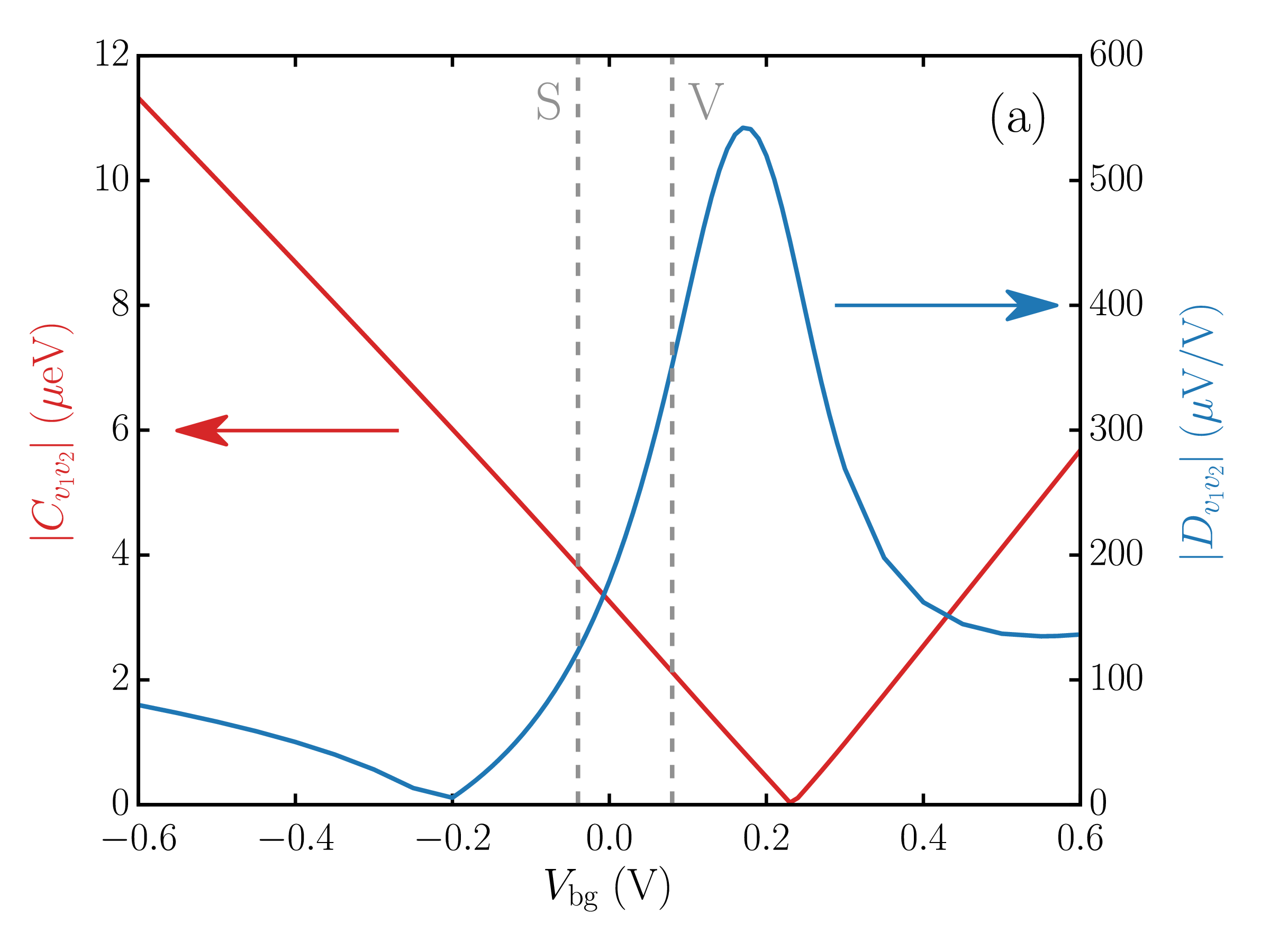} 
\includegraphics[width=.90\columnwidth]{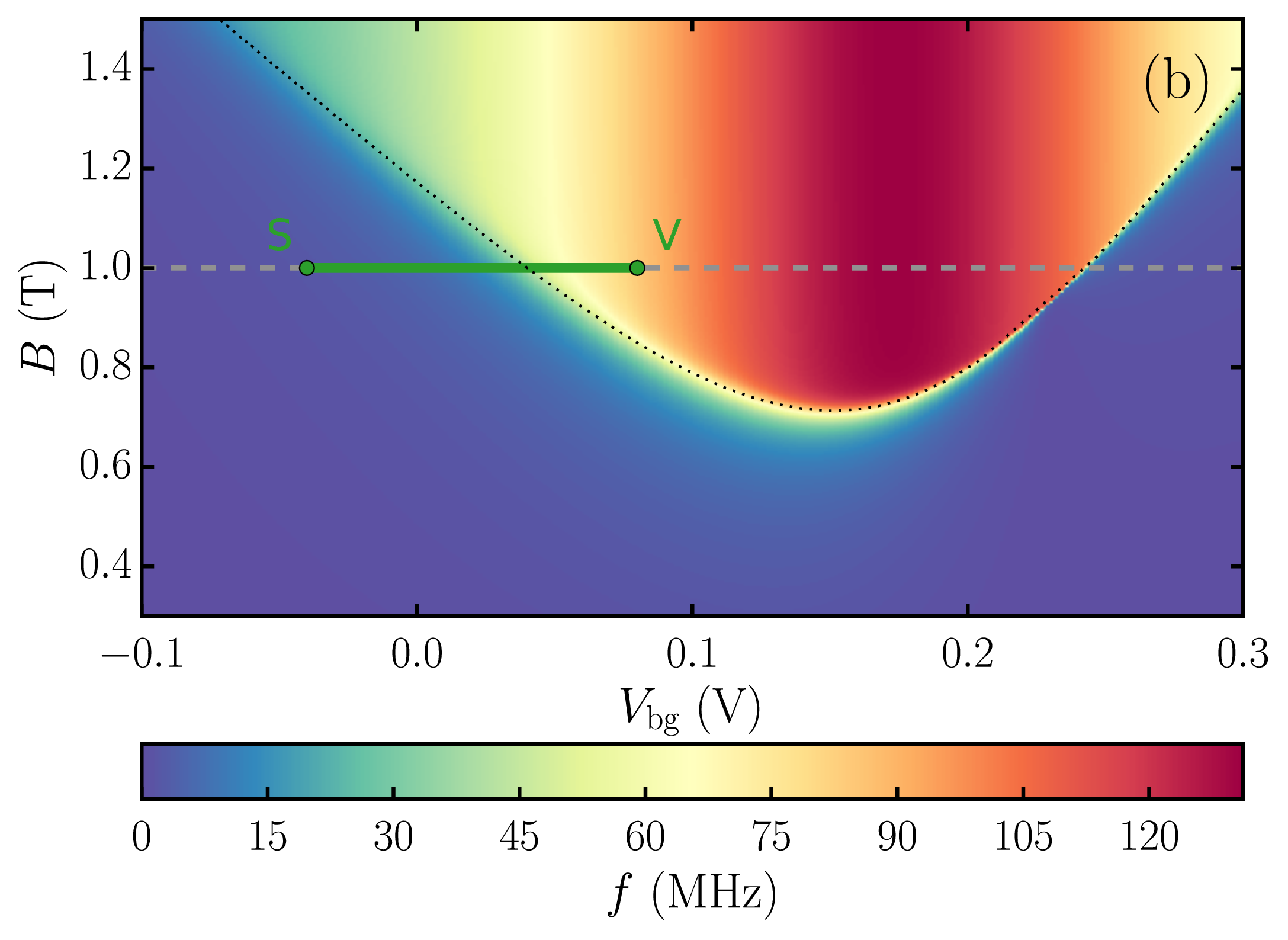} 
\caption{(a) $|C_{v_1v_2}|$ and $|D_{v_1v_2}|$ as a function of $V_{\rm bg}$. (b) Map of the Rabi frequency as a function of the magnetic field and $V_{\rm bg}$. The dotted black line is the anti-crossing condition $E_Z=g\mu_B B=\Delta(V_{\rm bg})$. A cut along the dashed gray line can be found in the Supporting Information. $V_{\rm fg}=0.1$ V and $\vec{B}\parallel y$ in all plots.}
\label{figRABIMAP}
\end{figure}

\begin{figure}
\includegraphics[width=.90\columnwidth]{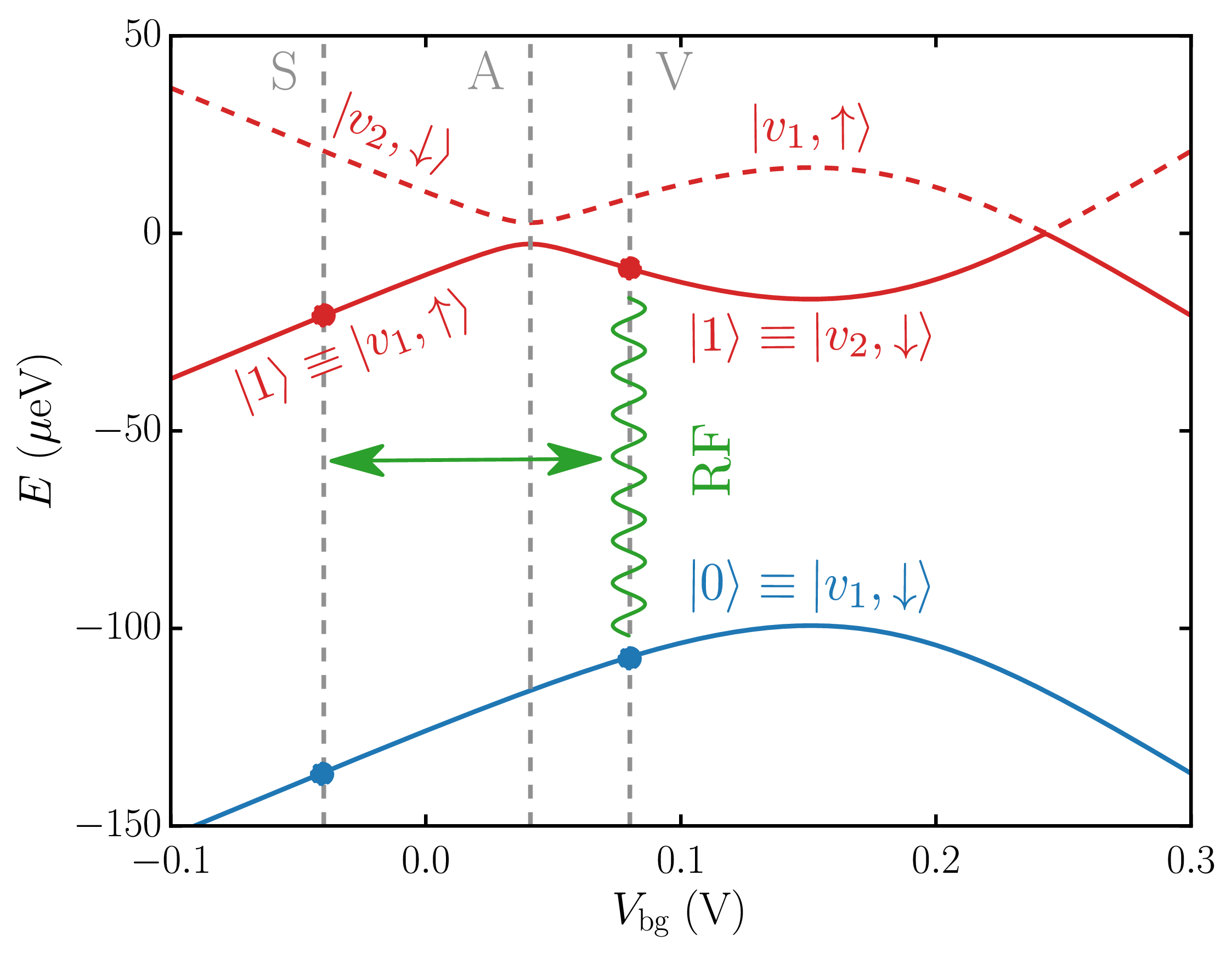} 
\includegraphics[width=.90\columnwidth]{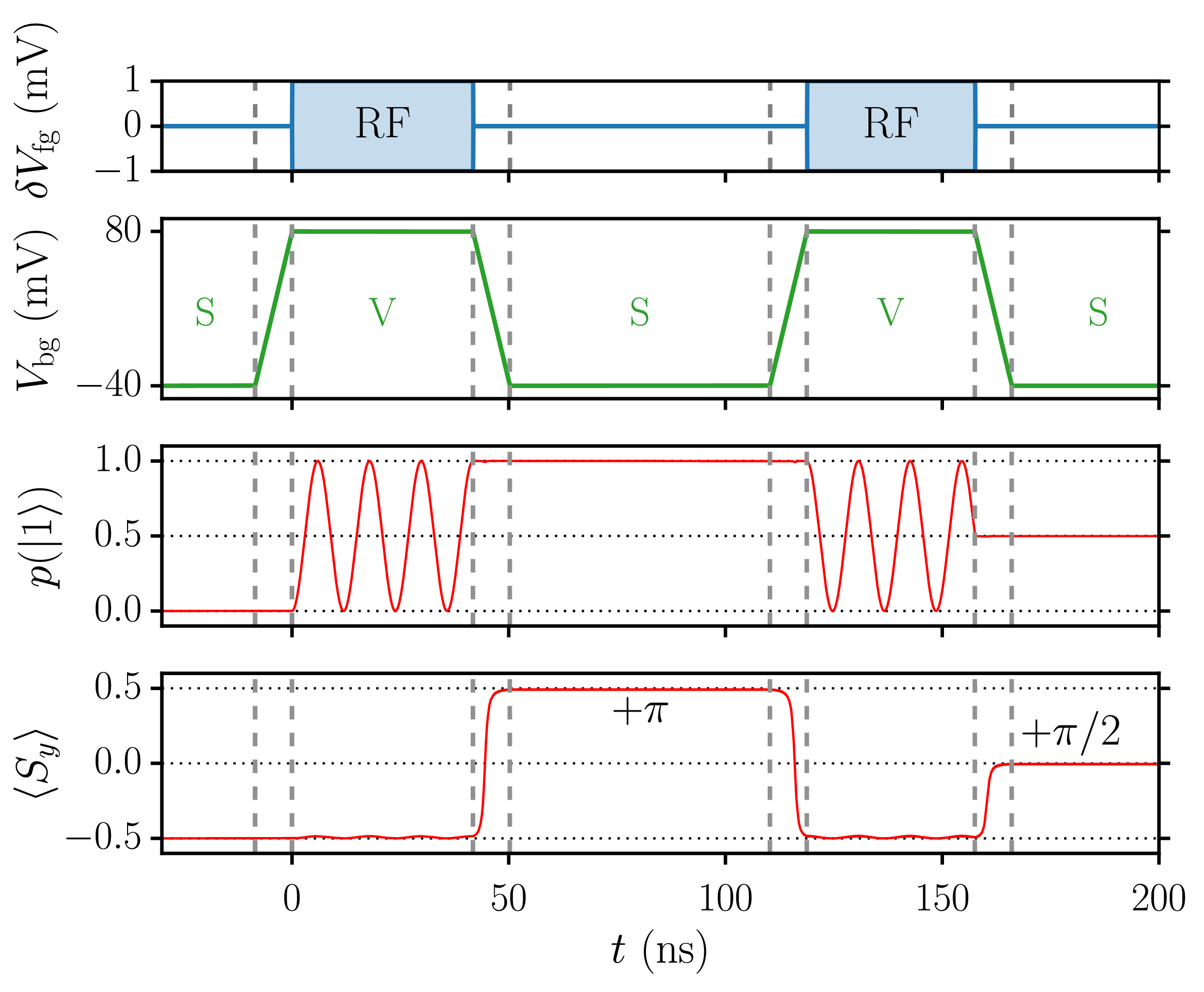} 
\caption{(top panel) Energy levels of the silicon QD as a function of $V_{\rm bg}$ ($V_{\rm fg}=0.1$ V, $B=1$ T along $y$, same colors as in Fig. \ref{figSV}). (bottom panels) Time series for spin manipulations, monitored by the probability $p(\ket{1})$ to be in the $\ket{1}$ state and by the average spin $\langle S_y\rangle$. Starting from the $\ket{0}\equiv\ket{v_1,\downarrow}$ state at point S in Fig. \ref{figRABIMAP}, the qubit is pulsed to point V by the back gate and a RF signal with frequency $\nu=23.66$ GHz on the front gate drives rotations between $\ket{0}$ and $\ket{1}\sim\ket{v_2,\downarrow}$; once the RF signal is switched off the qubit is brought back to point S where $\ket{1}\sim\ket{v_1,\uparrow}$ in order to complete the spin rotation.}
\label{figCHRONOGRAM}
\end{figure}

In order to illustrate this, we have performed tight-binding (TB) calculations using the $sp^3d^5s^*$ model of Ref. \onlinecite{Niquet09}. This model accounts for valley and spin-orbit coupling at the atomistic level. The potential in the device is first computed with a finite volumes Poisson solver, then the eigenstates of the TB Hamiltonian in this potential are calculated with a Jacobi-Davidson eigensolver. The Rabi frequencies are finally obtained from Eq. (\ref{eqrabid}), and spin manipulations are simulated with a time-dependent Schr\"odinger-Poisson solver in the basis of the 128 lowest-lying conduction band states of the QD. The atomistic segment of the device is 80 nm long and contains around $1\,120\,000$ atoms. The dangling bonds at the surface of the channel are saturated with pseudo-hydrogen atoms.

We first consider an ``ideal'' device without surface roughness disorder. The valley splitting $\Delta$ is plotted as a function of the back gate voltage $V_{\rm bg}$ at fixed front gate voltage $V_{\rm fg}=0.1$ V in Fig. \ref{figDELTA}. $\Delta$ decreases linearly with increasing $V_{\rm bg}$, then reaches a minimum in the 80 $\mu$eV range, and finally increases linearly again. During the back gate voltage sweep, the wave function of the electron moves from the top (negative $V_{\rm bg}$) to the bottom (positive $V_{\rm bg}$) interface, but remains confined under the top gate. The valley splitting increases when the wavefunction is further squeezed at one of the two interfaces by the vertical electric field, and is minimal when the electron is centered between the two interfaces. Although our model for the surface is simplified, the existing experimental data suggest that small valley splittings can indeed be achieved in SOI devices.\cite{Corna17} Also, test calculations made with the model of Ref. \onlinecite{Kim11} for the Si/SiO$_2$ interfaces show exactly the same trends.

$|C_{v_1v_2}|$ and $|D_{v_1v_2}|$ are plotted as a function of $V_{\rm bg}$ in Fig. \ref{figRABIMAP}a. They are little dependent on the magnitude of the magnetic field $\vec{B}\parallel y$. $C_{v_1v_2}\simeq 0$ just above $V_{\rm bg}^{\rm min}$ because the electron wave function, almost perfectly centered between the two gates, shows two additional horizontal $(xy)$ and vertical $(xz)$ quasi-symmetry planes.\cite{Corna17} $|D_{v_1v_2}|$ is, on the other hand, maximum near $V_{\rm bg}^{\rm min}$ because deconfinement in the $(yz)$ plane enhances coupling to the $z$ component of the RF electric field.

The calculated Rabi frequency is plotted as a function of $B$ and $V_{\rm bg}$ in Fig. \ref{figRABIMAP}b. The Rabi frequency is sizable within a hyperbolic-like shape whose edges are defined by the anti-crossing condition $E_Z=g\mu_B B=\Delta(V_{\rm bg})$. Indeed, for a given magnetic field $B$, there are typically zero or two back gate voltages that meet this condition (see Fig. \ref{figDELTA} and dotted line in Fig. \ref{figRABIMAP}b). The qubit goes in the valley regime inside the hyperbolic-like shape, and in the spin regime outside. The calculated Rabi frequency reaches values as large as 120 MHz near $V_{\rm bg}=V_{\rm bg}^{\rm min}$ where $|D_{v_1v_2}|$ is maximum. 

We can now design an electrical manipulation scheme taking advantage of Fig. \ref{figRABIMAP}. We set $B=1$ T along $y$ and bias the qubit along the line from point point S ($V_{\rm bg}=-0.04$ V) to  point V ($V_{\rm bg}=0.08$ V). At point V, the qubit is indeed in the valley regime and can be efficiently manipulated by the front gate (Rabi frequency $f\sim 80$ MHz). On the opposite, the qubit is in the spin regime at reference point S; the Rabi frequency is almost zero but the qubit is presumably much more robust to inelastic relaxation and decoherence than at point V. The energy levels along [SV] are plotted in the top panel of Fig. \ref{figCHRONOGRAM}.

The manipulation protocol is illustrated in the bottom panels of Fig. \ref{figCHRONOGRAM}, which represent the probability to be in the $\ket{1}$ state and the expectation value $\langle S_y\rangle$ of the spin along $y$ as a function of time during a $\pi$ and a $\pi/2$ rotation. The qubit is prepared in the $\ket{0}=\ket{v_1,\downarrow}$ state at point S, then switched to point V for manipulation. A RF pulse is applied on the front gate in order to drive a $\pi$ rotation, and the qubit is finally moved back to point S. The sequence is repeated for a subsequent $\pi/2$ rotation. Note that the system undergoes Rabi oscillations between states $\ket{0}\equiv\ket{v_1,\downarrow}$ and $\ket{1}\sim\ket{v_2,\downarrow}$ at point V; therefore, $\langle S_y\rangle$ remains almost constant at point V on Fig. \ref{figCHRONOGRAM}. However, at point S, $\ket{1}\equiv\ket{v_1,\uparrow}$, so that the spin rotations are completed by SOC on the way back from V to S.\cite{noteDissociate} It is important to sweep between S and V adiabatically enough so that the system remains on the lower branch $E_-$ of the anti-crossing and does not couple to the upper branch $E_+$ (which would result into a mixed spin/valley rotation back at point S). The slew rate on $V_{\rm bg}$ is primarily limited by the gap between $E_-$ and $E_+$ at the anti-crossing point A,\cite{Zener32} $E_{\rm SO}=2|C_{v_1v_2}|$. Here, $|C_{v_1v_2}|=2.7$ $\mu$eV is sufficiently large to achieve adiabatic switching within $<10$ ns. The possibility to drive arbitrary rotations is further demonstrated in the Supporting Information.

In order to assess spin coherence at points S and V, we have computed the relaxation time $T_1$ due to phonons and Johnson-Nyquist (JN) noise. We follow Refs. \onlinecite{Tahan14,Huang14} and assume a $2k\Omega$ series resistance on the front gate. We find that the operation of the qubit is limited by JN relaxation, with $T_1=64.6$ ms at point S, and $T_1=56.4$ $\mu$s at point V. As expected, the lifetimes are much longer in the spin than in the valley qubit regime, which is the rationale for this manipulation protocol. In the valley regime, $T_2^*$ might be strongly limited by the $1/f$ noise;\cite{Paladino14} we point out, though, that there is a sweet spot near $V_{\rm bg}=V_{\rm bg}^{\rm min}$, where the sensitivity of the valley splitting to gate and charge noise is minimal. More details about the models for $T_1$ and $T_2^*$ can be found in the Supporting Information.

We have also investigated the effects of surface roughness disorder on the Rabi frequencies (see Supporting Information). Surface roughness disorder reduces the valley splitting and is responsible for significant device-to-device variability. However, the valley-splitting $\Delta$ shows a minimum in the $\simeq 20-50$ $\mu$eV range near the same $V_{\rm bg}^{\rm min}$ in most devices, making the above manipulation protocol still possible with a proper calibration of each qubit. The Rabi frequencies are smaller because surface roughness reduces $|D_{v_1v_2}|$,\cite{Culcer10} yet they remain significant (typically $>20$ MHz).

To conclude, we have demonstrated that the mixing between the spin and valley degrees of freedom in a silicon qubit can be controlled by a suitable engineering of the electric field. Thanks to the weak, but sizable spin-orbit coupling in the conduction band, the qubit can be continuously switched from a ``spin'' to a mixed ``spin-valley'' and eventually a ``valley'' mode by the action on the gates. In the spin-valley and valley modes, Rabi oscillations can be driven by radio-frequency signals on the gates, allowing for all-electrical manipulation schemes. In the pure spin mode, the qubit is not electrically addressable but is much more robust to inelastic relaxation and decoherence. A spin qubit may hence be switched to the valley mode for electrical manipulation then back to the spin mode in order to benefit from the long spin coherence times afforded by silicon. These findings open new perspectives for the development of efficient and scalable spin qubits on silicon. They also confirm that the effects of spin-orbit coupling in the conduction band of silicon are far from negligible, and can even be tailored for practical applications. 

We thank Louis Hutin, Beno\^it Bertrand and Silvano de Franceschi for fruitful discussions. This work was supported by the European Union's Horizon 2020 research and innovation program under grant agreement No 688539 MOSQUITO. Part of the calculations were run on the TGCC/Curie and CINECA/Marconi machines using allocations from GENCI and PRACE.

%

\pagebreak

\widetext
\begin{center}
\textbf{\large Supporting Information for ``All-electrical manipulation of silicon spin qubits with tunable spin-valley mixing''}
\end{center}
\setcounter{equation}{0}
\setcounter{figure}{0}
\setcounter{table}{0}
\setcounter{page}{1}
\makeatletter
\renewcommand{\theequation}{S\arabic{equation}}
\renewcommand{\thefigure}{S\arabic{figure}}
\renewcommand{\bibnumfmt}[1]{[S#1]}
\renewcommand{\citenumfont}[1]{S#1}


\FloatBarrier

In this Supporting Information, we provide a line cut on on Fig. 4 of the main text (section \ref{sectionLC}), then discuss the nature of the rotations performed with the present manipulation protocol (section \ref{sectionRotations}), and finally the effects of surface roughness (section \ref{sectionSR}) and the calculation of $T_1/T_2^*$ (section \ref{sectionTs}).

\section{Line cut on Fig. 4 of the main text.}
\label{sectionLC}

\begin{figure}
\includegraphics[width=.475\columnwidth]{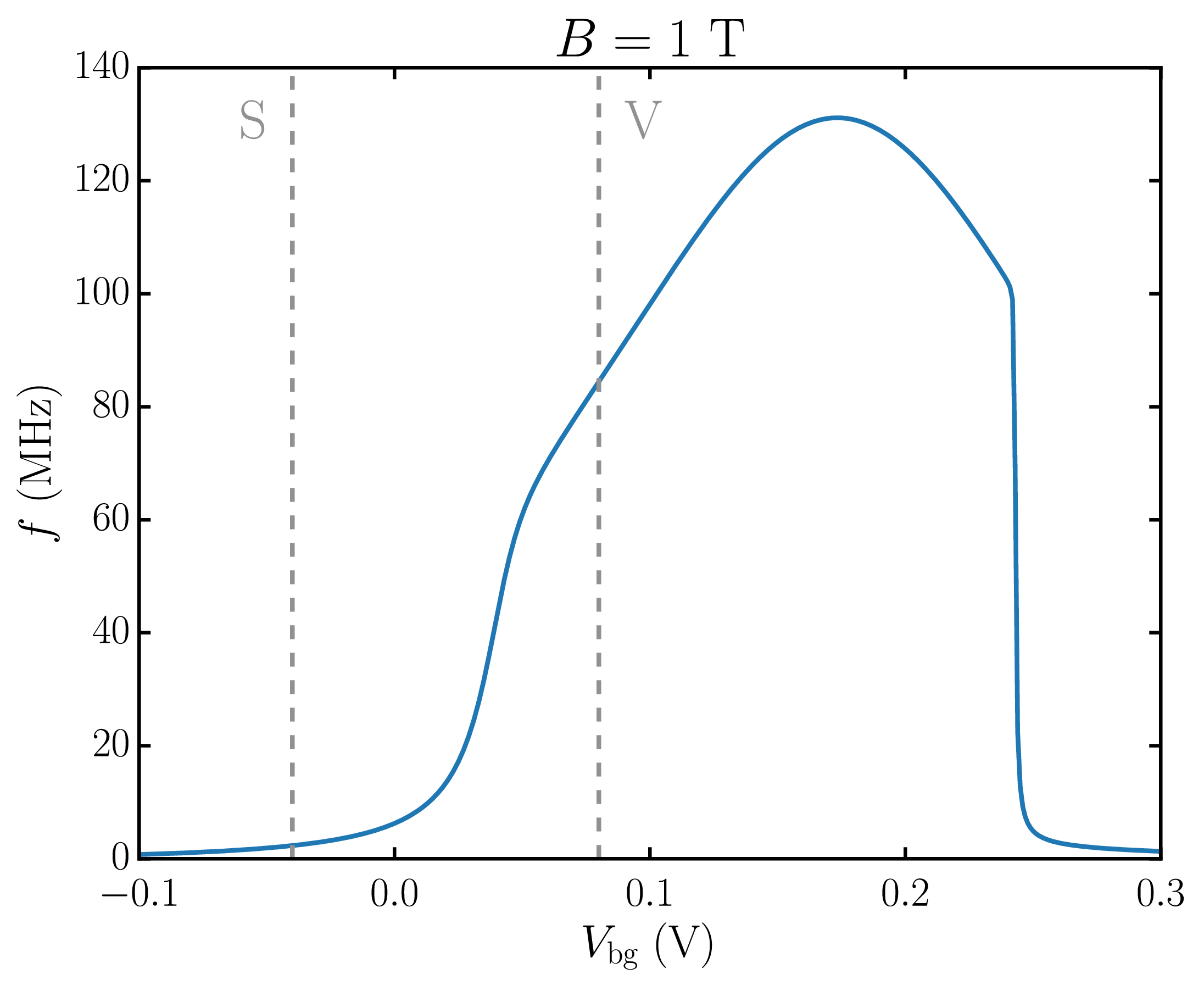} 
\caption{Rabi frequency as a function of the back gate voltage $V_{\rm bg}$ at magnetic field $B=1$ T along $y=[1\bar{1}0]$ (cut along the horizontal dashed gray line in Fig. 4b of the main text).}
\label{figRABIMAP3}
\end{figure}

The Rabi frequency is plotted in Fig. \ref{figRABIMAP3} along the horizontal dashed line of Fig. 4b, main text. The width of the transitions from the spin to the valley qubit regimes is controlled by the SOC matrix element $C_{v_1v_2}$, while the Rabi frequency in the valley qubit regime is essentially set by the gate coupling matrix element $D_{v_1v_2}$ (Fig. 4a). 

\section{Nature of the rotations.}
\label{sectionRotations}

\begin{figure}
\includegraphics[width=.475\columnwidth]{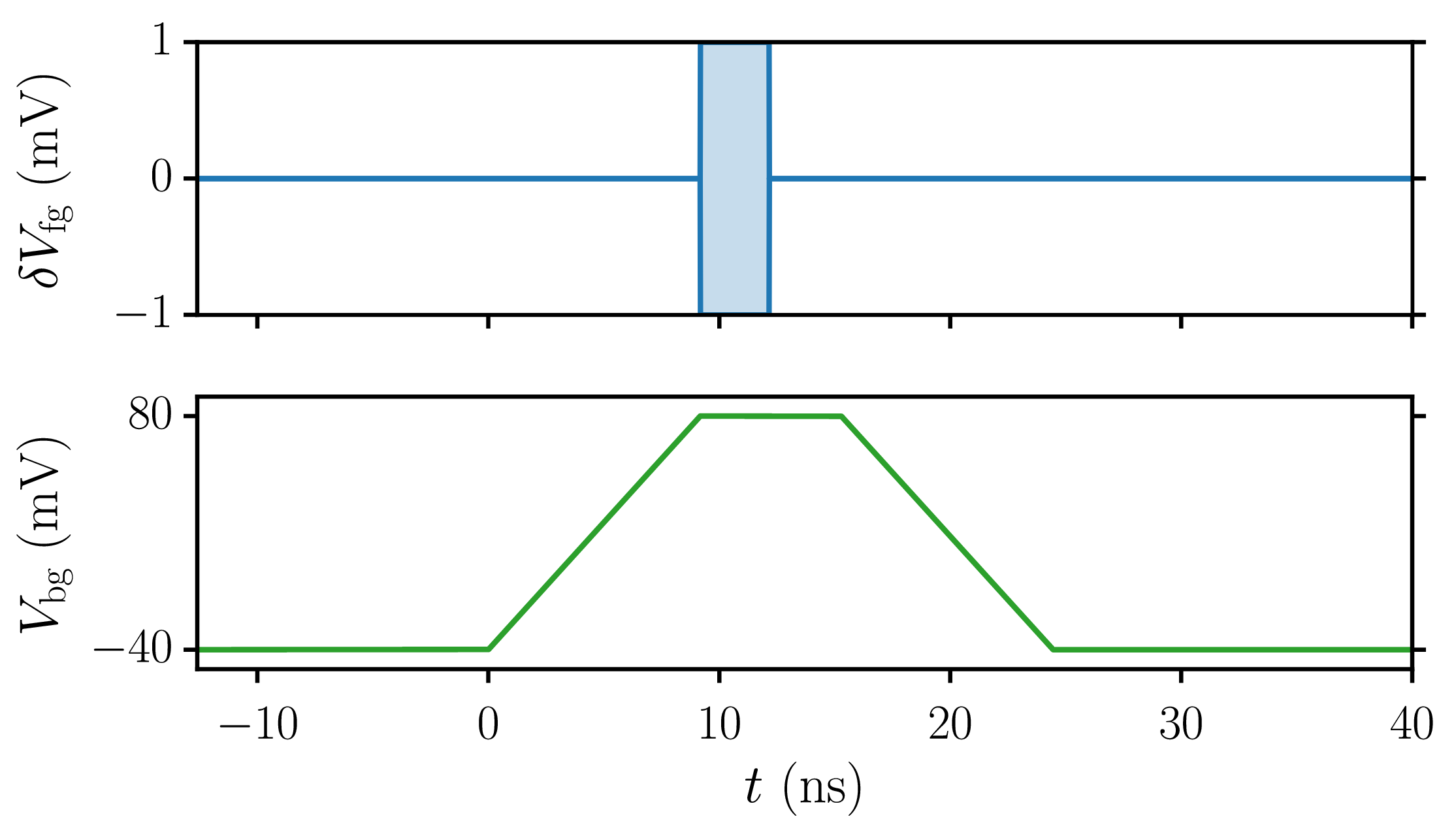} \\
\includegraphics[width=.475\columnwidth]{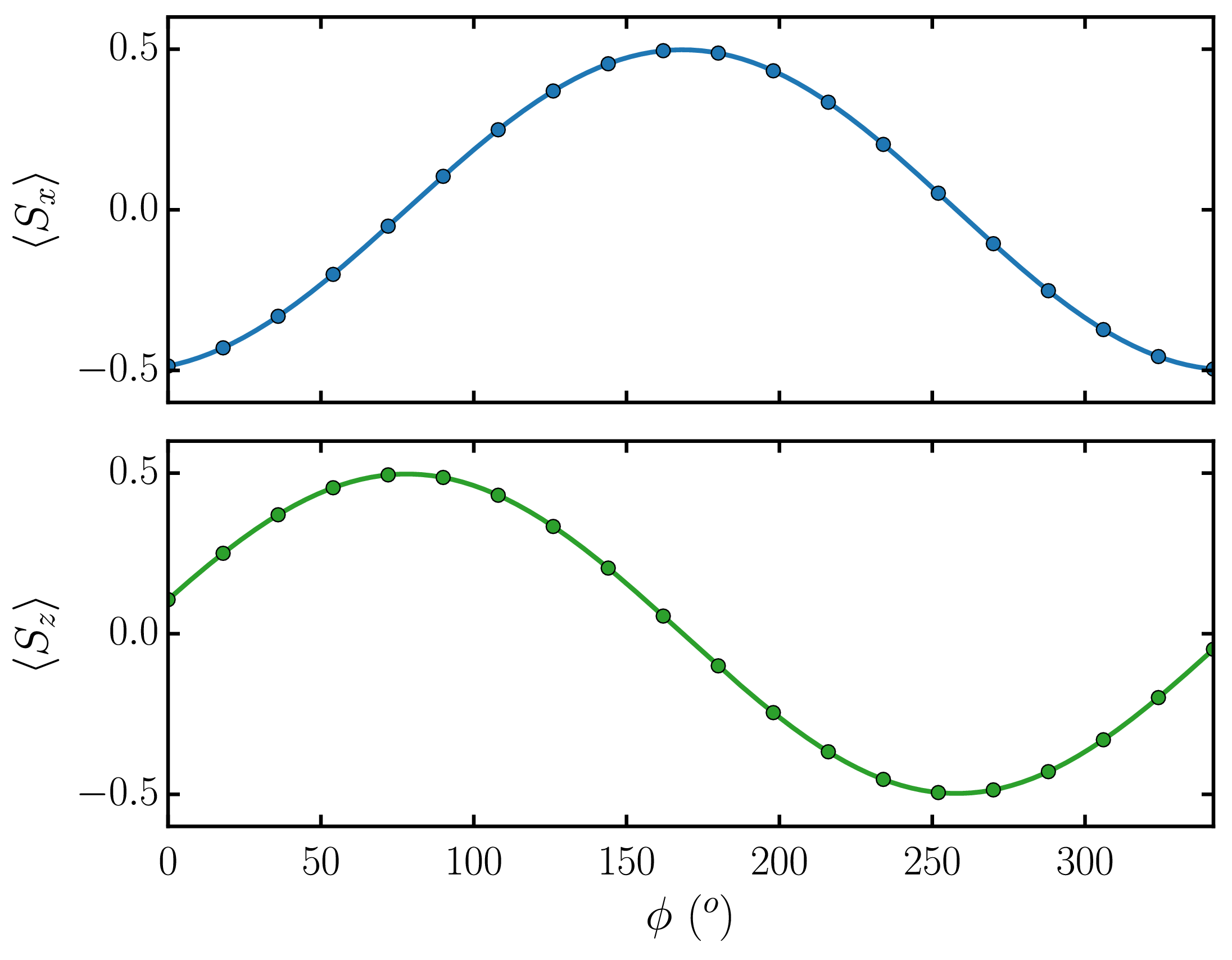} 
\caption{Time series for a $\pi/2$ rotation from the $\ket{v_1,\downarrow}$ state, and expectation value of $S_x$ and $S_z$ in the rotating Bloch sphere at S after that $\pi/2$ rotation, as a function of the phase $\phi$ of the driving RF signal (same system as in the main text). The magnetic field $\vec{B}$ is oriented along $y$.}
\label{figROTATIONS}
\end{figure}

During the manipulation sequence, the phase of the qubit drifts on the way from S to V then from V to S, as well as during the rotation at V, since the precession frequencies are slightly different at the S and V points. Let us, therefore, introduce the time-dependent states $\ket{0}(t)=\ket{v_1,\downarrow}e^{+i\omega_{\rm S}t/2}$ and $\ket{1}(t)=\ket{v_1,\uparrow}e^{-i\omega_{\rm S}t/2}$, where $\omega_{\rm S}/(2\pi)$ is the precession frequency at point S. The projections of the qubit state on $\ket{0}(t)$ and $\ket{1}(t)$ define its representation in the rotating Bloch sphere at point S.

The transformation matrix $T$ for the manipulation sequence reads in the $\{\ket{0}(t),\ket{1}(t)\}$ basis set:
\begin{equation}
T=R_Z(\Delta\varphi_{\rm VS})R_Z(\Delta\varphi_{\rm V})R_{XY}(\alpha, \varphi)R_Z(\Delta\varphi_{\rm SV})\,,
\end{equation}
where $R_Z(\alpha)$ is the matrix of a rotation of angle $\alpha$ around the polar axis $\vec{Z}$ of the Bloch sphere:
\begin{equation}
R_Z(\alpha)=
\begin{pmatrix}{}
e^{i\alpha/2} & 0 \\
0 & e^{-i\alpha/2}
\end{pmatrix}\,,
\end{equation} 
and $R_{XY}(\alpha, \varphi)$ is the matrix of a rotation of angle $\alpha$ around  $\vec{U}=\cos\varphi\vec{X}+\sin\varphi\vec{Y}$:
\begin{equation}
R_{XY}(\alpha, \varphi)=
\begin{pmatrix}{}
\cos(\alpha/2) & -i\sin(\alpha/2)e^{i\varphi} \\
-i\sin(\alpha/2)e^{-i\varphi} & \cos(\alpha/2)
\end{pmatrix}\,.
\end{equation} 
$\Delta\varphi_{\rm SV}$, $\Delta\varphi_{\rm V}$ and $\Delta\varphi_{\rm VS}$ are the phase shifts accumulated on the way from S to V, at the V point, and back from V to S. $\Delta\varphi_{\rm SV}$ and $\Delta\varphi_{\rm VS}$ depend on the back gate voltage ramps, while $\Delta\varphi_{\rm V}=(\omega_{\rm V}-\omega_{\rm S})\tau_{\rm V}$, where $\omega_{\rm V}/(2\pi)$ and $\tau_{\rm V}$ are the precession frequency and the total time spent at point V, respectively. $\alpha$ is controlled by the duration $\tau_\alpha\le\tau_{\rm V}$ of the RF pulse at V. The axis of rotation, characterized by $\varphi$, can in principle be controlled by the phase of the RF signal, as demonstrated below.

The above sequence of rotations can be factorized as:
\begin{equation}
T=R_Z(\Delta\varphi_{\rm SV}+\Delta\varphi_{\rm V}+\Delta\varphi_{\rm VS})R_{XY}(\alpha,\varphi-\Delta\varphi_{\rm SV})\,.
\end{equation}
Namely, the net operation appears as a rotation around an axis of the equatorial plane of the Bloch sphere (as expected), followed by a rotation around $\vec{Z}$ that accounts for the total phase accumulated out of the S point. This phase must be accounted for when chaining rotations. It can be compensated by choosing $\tau_{\rm V}$ such that $\Delta\varphi_{\rm T}=\Delta\varphi_{\rm SV}+\Delta\varphi_{\rm V}(\tau_{\rm V})+\Delta\varphi_{\rm VS}=2n\pi$ irrespective of the rotation (typically, $\tau_{\rm V}$ must be greater than $\tau_\pi$ so that $\pi$ rotations can be accommodated within the manipulation window at V).

As an illustration, figure \ref{figROTATIONS} shows the expectation value of $S_x$ and $S_z$ in the rotating Bloch sphere after a $\pi/2$ rotation from the $\ket{v_1,\downarrow}$ state, as a function of the phase $\phi$ of the RF signal on the front gate [namely, $\delta V_{\rm fg}(t)\propto\sin(\omega_{\rm V} t+\phi)$]. The magnetic field $\vec{B}$ is parallel to $y$. This figure confirms that rotations can be driven around arbitrary axes of the equatorial plane of the rotating Bloch sphere by controlling the phase of the RF signal, as done in conventional ESR/EDSR experiments.

In this figure, the time $\tau_{\rm V}$ spent at the V point has been adjusted so that two successive $\pi/2$ rotations around the same axis result in a net $\pi$ rotation ($\Delta\varphi_{\rm T}=2n\pi$). Still, the phase $\phi$ of the second rotation must account for the mismatch in precession frequencies at S and V. For example, if the first rotation at time $t_0$ is driven by a RF signal $\delta V_{\rm fg}(t)\propto\sin[\omega_{\rm V} (t-t_0)+\phi)]$, the second rotation at time $t_1$ must be driven by a RF signal $\delta V_{\rm fg}(t)\propto\sin[\omega_{\rm V} (t-t_0)+\phi+(\omega_{\rm S}-\omega_{\rm V})(t_1-t_0)]$.

\section{Effects of surface roughness.}
\label{sectionSR}

\begin{figure}
\includegraphics[width=.32\columnwidth]{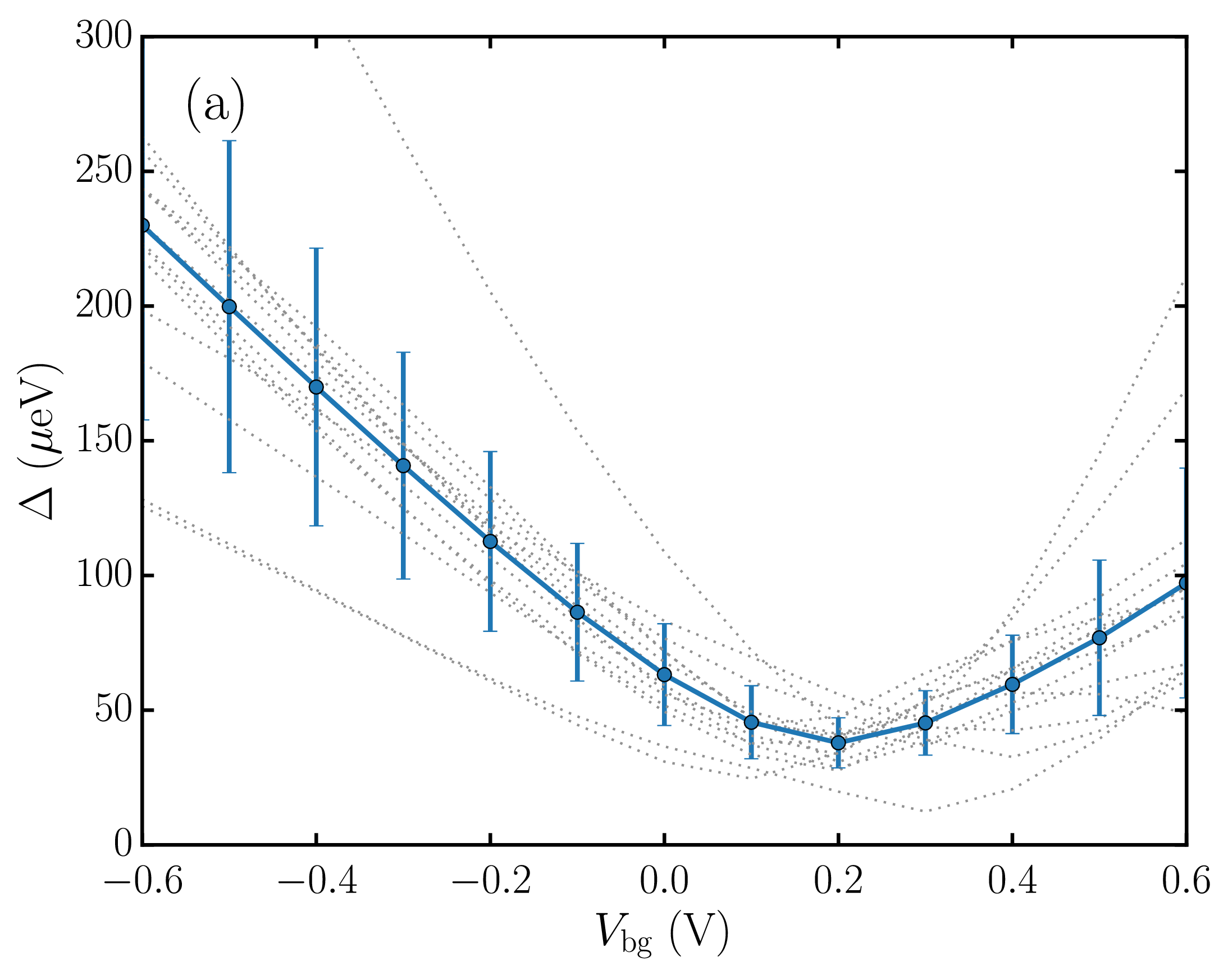} 
\includegraphics[width=.32\columnwidth]{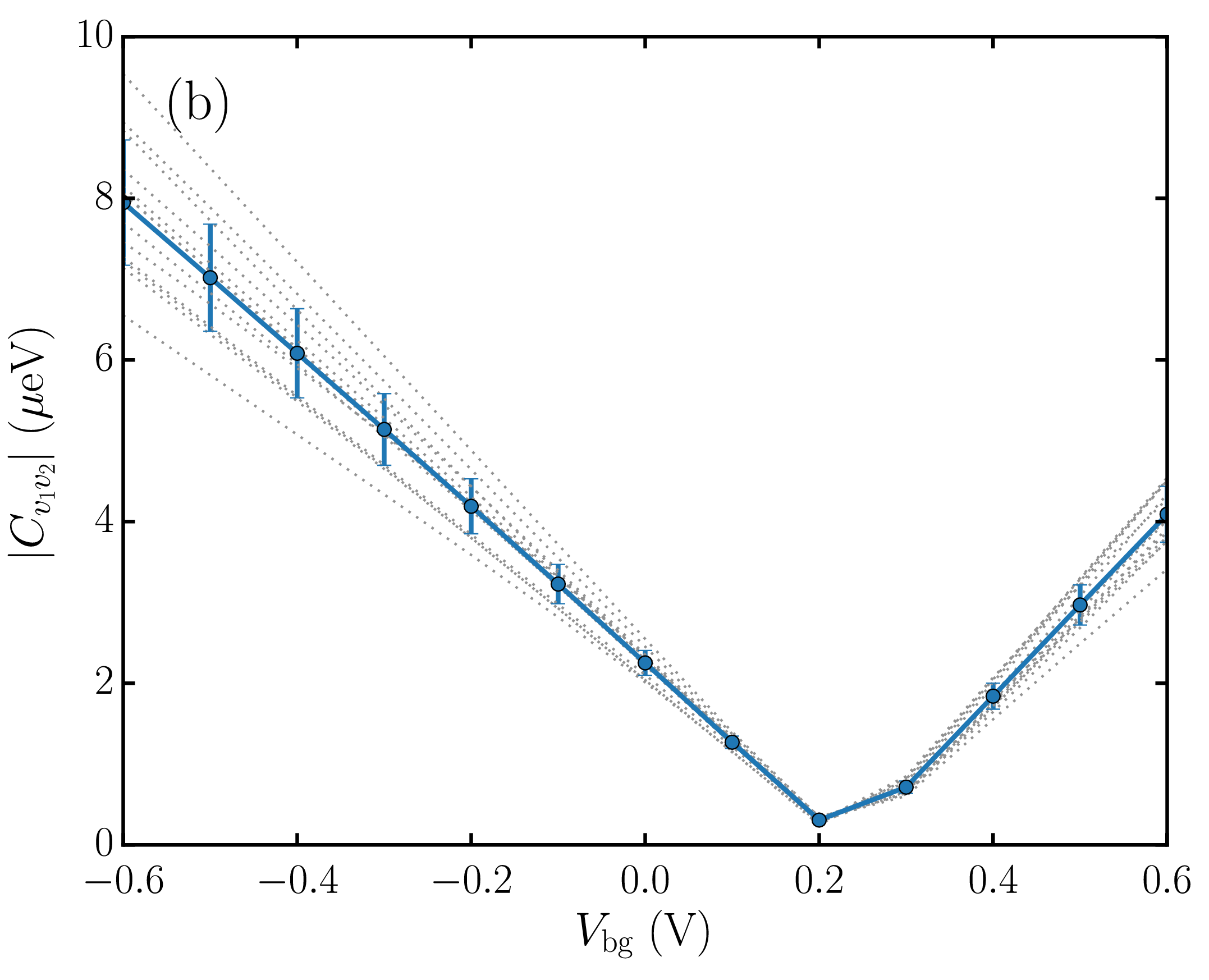} 
\includegraphics[width=.32\columnwidth]{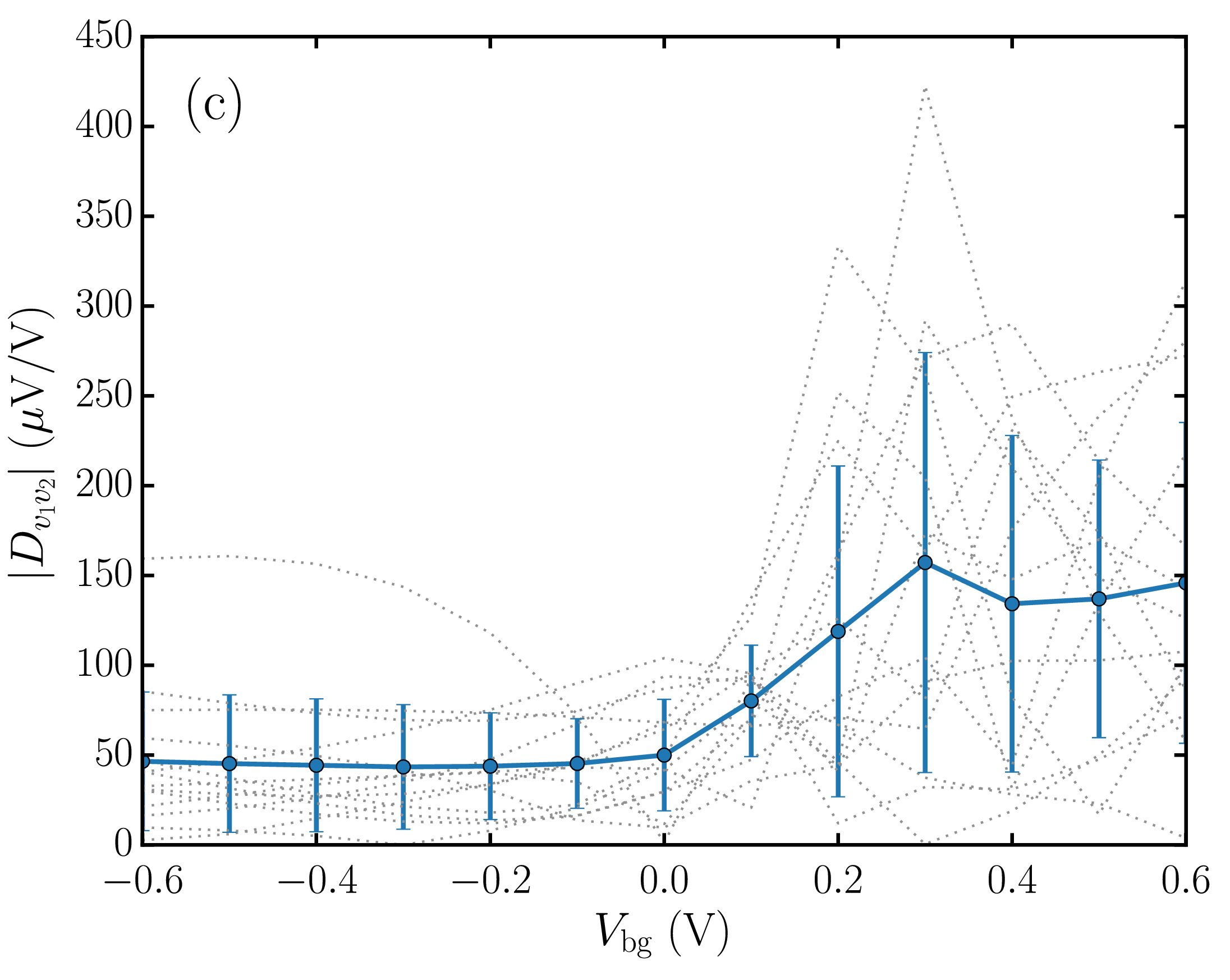} 
\caption{(a) Valley splitting $\Delta$ as a function of the back gate voltage $V_{\rm bg}$, for different realizations of the surface roughness disorder (dotted gray lines). The average and standard deviation are plotted as the blue line and error bars. (b) SOC matrix element $C_{v_1v_2}$ as a function of the back gate voltage $V_{\rm bg}$, for different realizations of the disorder (c) Gate coupling matrix element $D_{v_1v_2}$ as a function of the back gate voltage $V_{\rm bg}$, for different realizations of the disorder. $V_{\rm fg}=0.1$ V in all plots.}
\label{figDELTA2}
\end{figure}

In order to assess the robustness and variability of the results, we have introduced surface roughness (SR) disorder in the simulations. The SR profiles are generated from a Gaussian spectral density with rms $\Delta_{\rm SR}=0.4$ nm and correlation length $\Lambda_{\rm SR}=1.5$ nm.\cite{Goodnick85} $\Delta_{\rm SR}$ lies in the upper range of the values compatible with the carrier mobilities measured in similar devices at room temperature.\cite{Bourdet16} The SR profiles are, therefore, pretty aggressive. Surface roughness might be mitigated with suitable annealing techniques.\cite{Dornel07}

The valley-splitting $\Delta$ is plotted as a function of the back gate voltage $V_{\rm bg}$ in Fig. \ref{figDELTA2}a for different realizations of the disorder. Although the slope of $\Delta(V_{\rm bg})$ shows significant variability on both front and back interfaces, most curves show a minimum $\Delta^{\rm min}$ in the $25-55$ $\mu$eV range. $\Delta^{\rm min}$ is smaller with SR ($\Delta^{\rm min}=83$ $\mu$eV without), because roughness averages out part of the valley interactions.\cite{Culcer10} This brings the manipulation frequency in the valley qubit regime down to the $\sim 5-15$ GHz range, which is easily accessible with standard RF circuitry.

\begin{figure}
\includegraphics[width=.475\columnwidth]{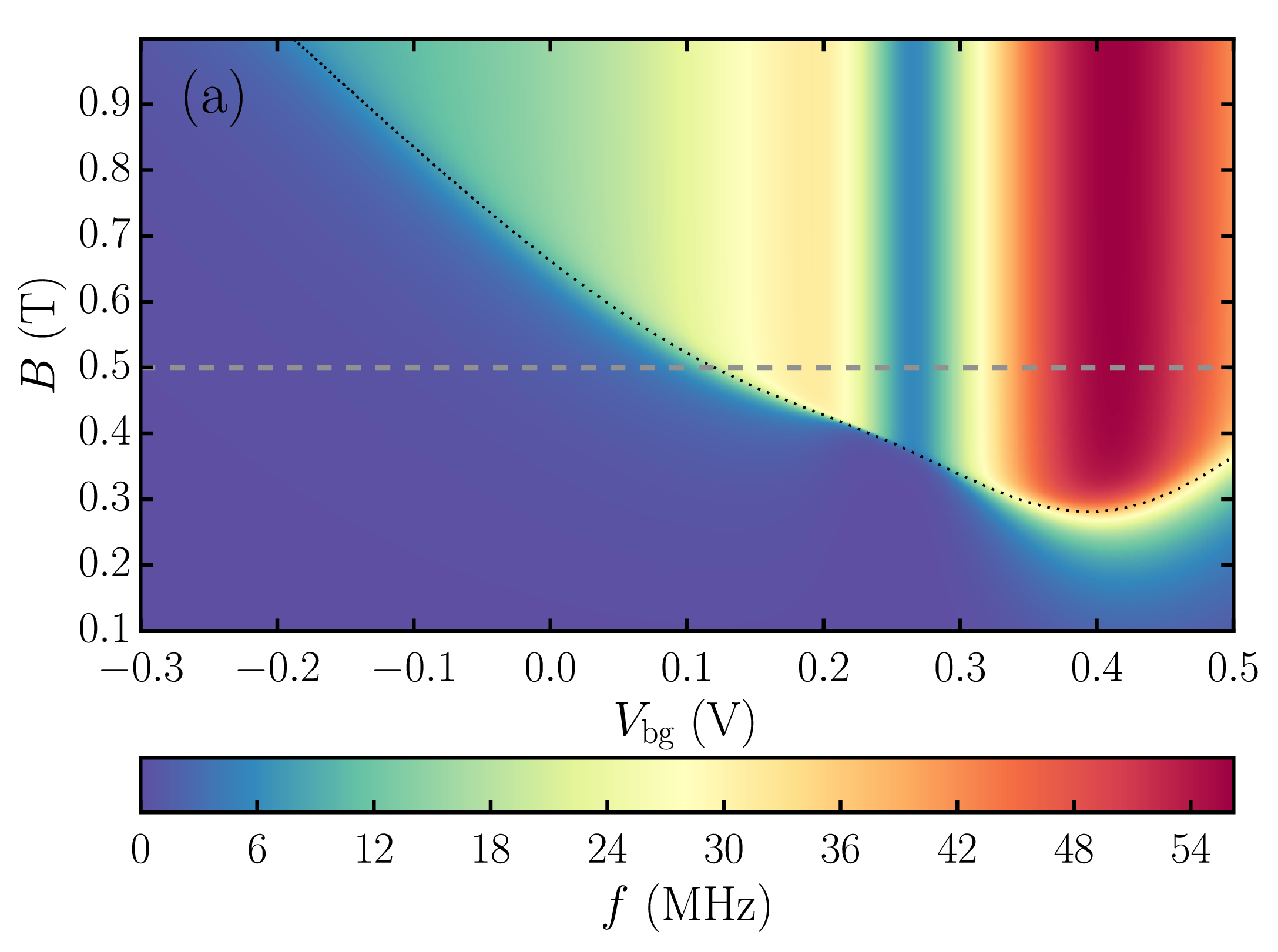} 
\includegraphics[width=.475\columnwidth]{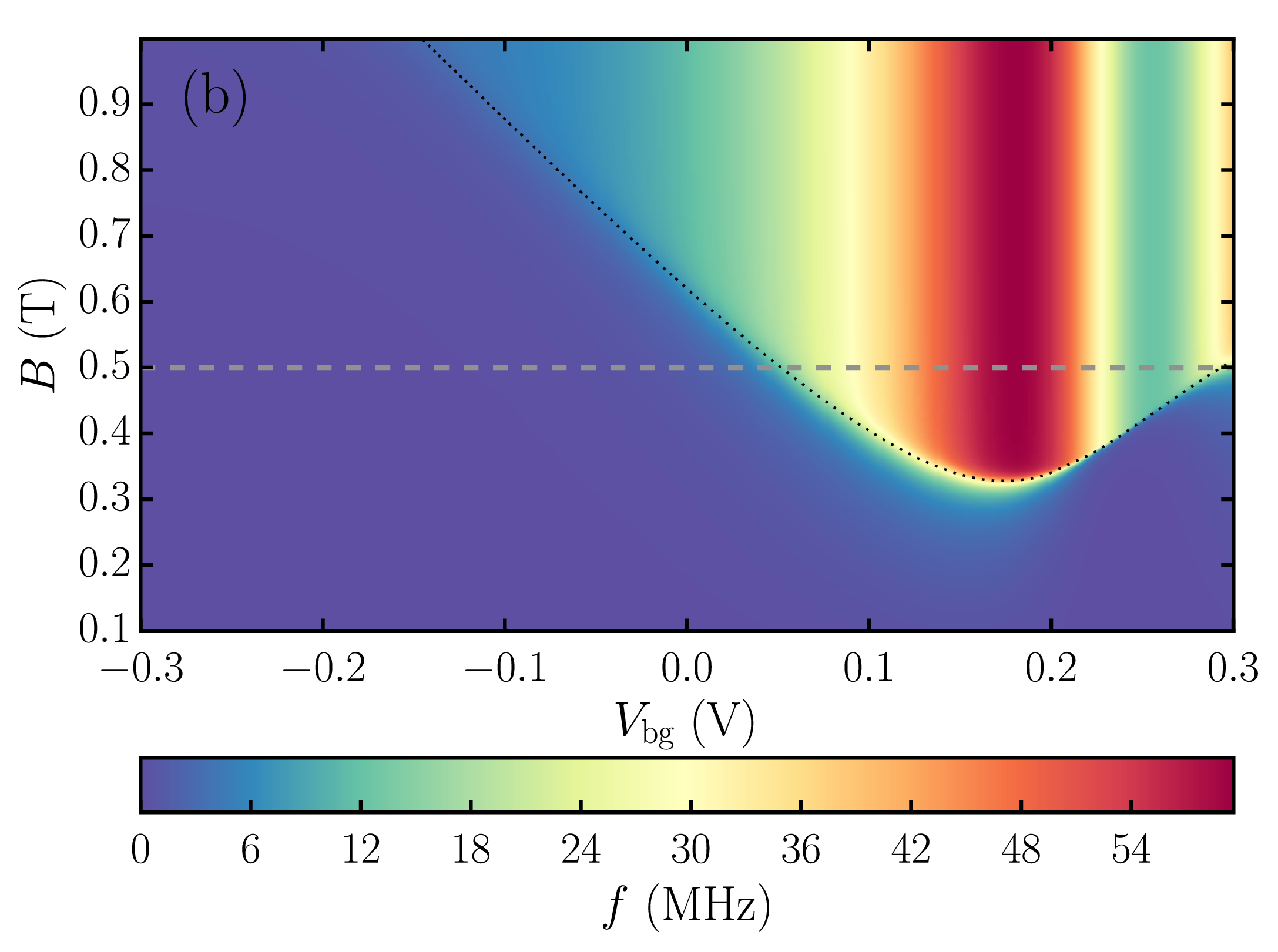} 
\includegraphics[width=.475\columnwidth]{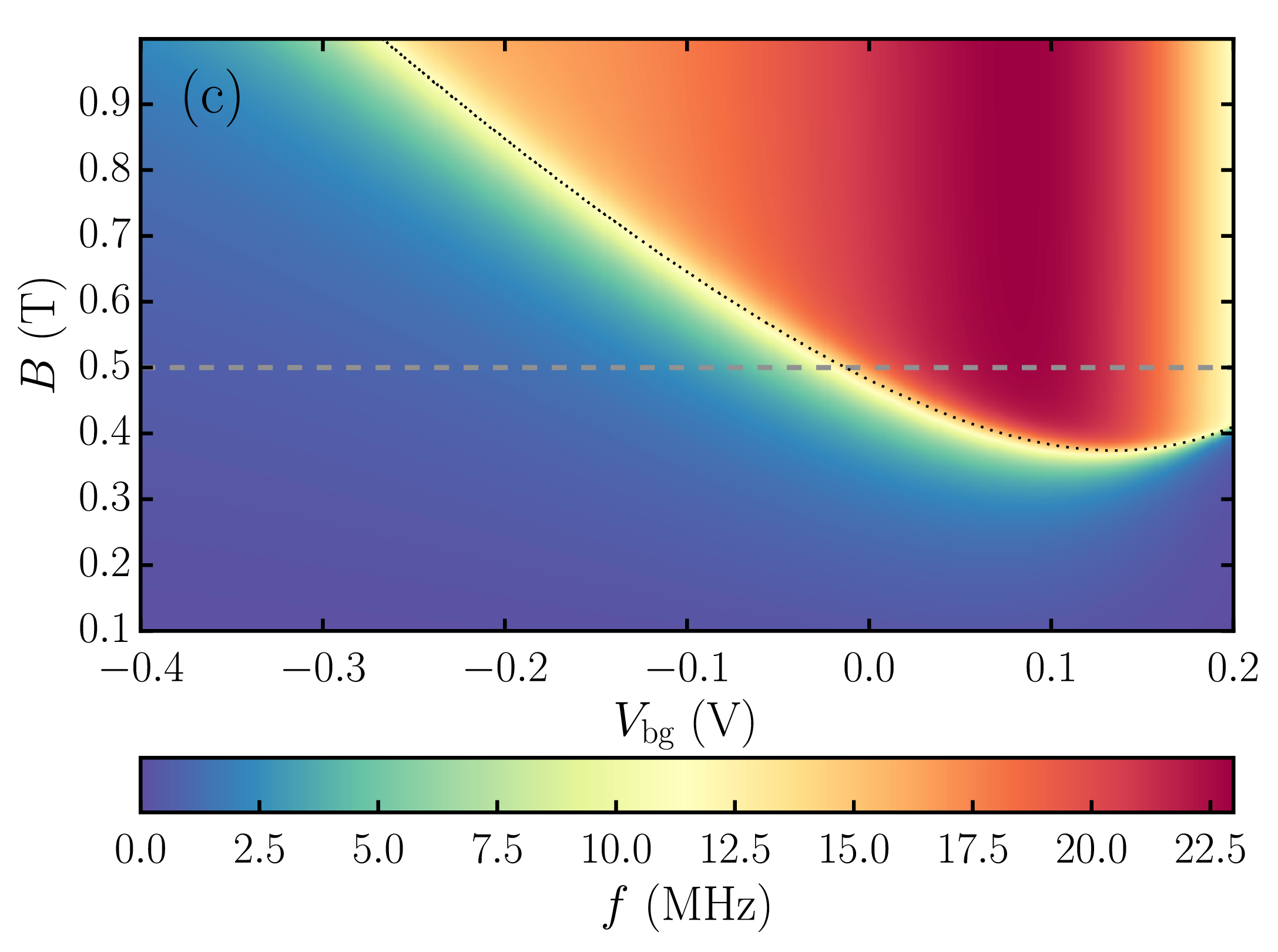} 
\includegraphics[width=.475\columnwidth]{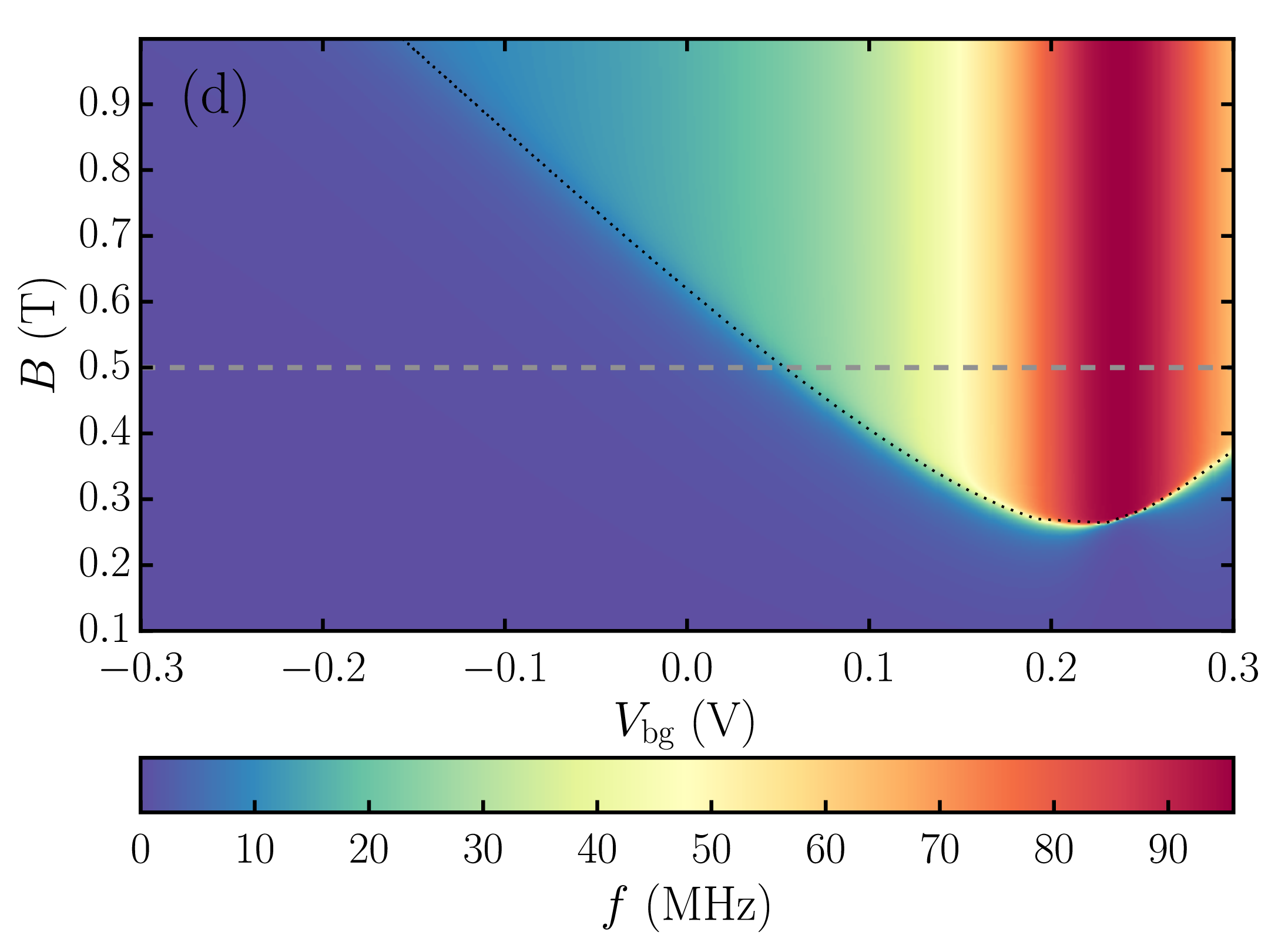} 
\caption{Map of the Rabi frequency as a function of the magnetic field and $V_{\rm bg}$, for different realizations of the disorder. The dotted black line is the anti-crossing condition $E_Z=g\mu_B B=\Delta(V_{\rm bg})$. The horizontal dashed line is a target magnetic field $B=0.5$ T for qubit operation. $V_{\rm fg}=0.1$ V and $\vec{B}\parallel y$ in all plots.}
\label{figRABIMAP2}
\end{figure}

The matrix elements $C_{v_1v_2}$ and $D_{v_1v_2}$ are plotted in Fig. \ref{figDELTA2}b and \ref{figDELTA2}c for different realizations of the disorder. They are, likewise, both decreased by the roughness. $C_{v_1v_2}$ shows little variability, while the shape and magnitude of $D_{v_1v_2}$ can be strongly dependent on the particular realization of the SR, especially near $V_{\rm bg}=V_{\rm bg}^{\rm min}$ due to the complex interference pattern between the top and down interfaces. This may lower the achievable Rabi frequencies, but does not, in general, preclude the proposed manipulation protocol at the price of a calibration of each qubit. This is illustrated in Fig. \ref{figRABIMAP2}, which shows maps of the Rabi frequency as a function of the magnetic field and $V_{\rm bg}$ for four different realizations of the disorder. Although some maps might show a more complex behavior than Fig. 4 of the main text, the qubit remains electrically addressable over a wide range of back gate voltages within the valley regime. The calculated Rabi frequencies typically reach a few tens of MHz, which is still very significant.

\section{Calculation of $T_1$ and $T_2^*$.}
\label{sectionTs}

We compute the relaxation rate due to the electron-phonon interactions in the electric dipole approximation.\cite{Tahan14,Huang14} The contribution from longitudinal phonons reads:
\begin{align}
T_{1,\rm l}^{-1}&=\frac{\omega_{01}^5}{2\pi\hbar\rho v_l^7}\left[\left(|X_{01}|^2+|Y_{01}|^2\right)\left(\frac{1}{3}\Xi_d^2+\frac{2}{15}\Xi_d\Xi_u+\frac{1}{35}\Xi_u^2\right)\right. \nonumber \\
&+\left.|Z_{01}|^2\left(\frac{1}{3}\Xi_d^2+\frac{2}{5}\Xi_d\Xi_u+\frac{1}{7}\Xi_u^2\right)\right]{\rm coth}\left(\frac{\hbar\omega_{01}}{2kT}\right)
\end{align}
while the contribution from transverse phonons reads:
\begin{equation}
T_{1,\rm t}^{-1}=\frac{\omega_{01}^5}{2\pi\hbar\rho v_t^7}\left[\left(|X_{01}|^2+|Y_{01}|^2\right)\frac{4}{105}\Xi_u^2+|Z_{01}|^2\frac{2}{35}\Xi_u^2\right]{\rm coth}\left(\frac{\hbar\omega_{01}}{2kT}\right)\,,
\end{equation}
where $\omega_{01}/(2\pi)$ is the qubit precession frequency, $X_{01}=\langle0|x|1\rangle$, $Y_{01}=\langle0|y|1\rangle$ and $Z_{01}=\langle0|z|1\rangle$ are the dipole matrix elements in the device axis set, $v_l=9000$ m/s and $v_t=5400$ m/s are the longitudinal and transverse sound velocities, $\Xi_d=1.0$ eV and $\Xi_u=8.6$ eV are the conduction band deformation potentials, $\rho=2329$ kg/m$^3$ is the mass density of silicon, and $T=100$ mK is the temperature.

We follow Refs. \onlinecite{Huang14}, \onlinecite{Clerk10} and \onlinecite{Paladino14} for Johnson-Nyquist noise. We assume a $R=2k\Omega$ series resistance on the front gate so that:
\begin{align}
T_{1,\rm jn}^{-1}&=\frac{4\pi}{\hbar}\frac{R}{R_0}|D_{01}|^2\hbar\omega_{01}{\rm coth}\left(\frac{\hbar\omega_{01}}{2kT}\right) \nonumber \\
T_{2,\rm jn}^{*-1}&=\frac{2\pi}{\hbar}\frac{R}{R_0}|D_{11}-D_{00}|^2kT\,,
\end{align}
where $R_0=h/e^2$, $D_{00}=\langle0|D|0\rangle$, $D_{11}=\langle1|D|1\rangle$ and $D_{01}=\langle0|D|1\rangle$.

The relevant data at the S and V points are given in Table \ref{TableT} for the device of the main text. As expected, $T_1$ and $T_2^*$ are much longer in the spin than in the valley regime due to the reduced sensitivity of spin qubits to electric fields. The operation of the qubit is limited by Johnson Nyquist noise, but the calculated $T_{1,\rm jn}$ remains orders of magnitude larger than the total manipulation time (a few tens of ns on Fig. 5, main text and on Fig. \ref{figROTATIONS}). The phonon-limited $T_1$'s are also much longer than measured in Ref. \onlinecite{Yang13} because the valley splittings and dipole matrix elements are smaller (in particular, $T_{1,\rm l}$ and $T_{1,\rm t}$ scale as $\Delta^{-5}$ in the valley regime). Practically, the coherence might be limited by various sources of $1/f$ noise\cite{Paladino14} (charge and gate noise, ...), which still need to be carefully characterized.

\begin{table}
{\setlength\doublerulesep{0.5pt}
\begin{tabular}{lrr}
\toprule[1pt]\midrule[0.5pt]
			        & S point               & V point \\
\midrule
$\hbar{\omega_{01}}$ ($\mu$eV)  & 115.3                 & 98.3 \\
$X_{01}$ (\AA)                  & 0.000                 & 0.001 \\  
$Y_{01}$ (\AA)                  & 0.005                 & 0.050 \\
$Z_{01}$ (\AA)                  & 0.011                 & 0.287 \\
$D_{01}$ ($\mu$V/V)             & 9.5                   & 348.9 \\    
$|D_{11}-D_{00}|$ ($\mu$V/V)    & 2.4                   & 607.8 \\      
$T_{1,\rm l}^{-1}$ (s$^{-1}$)   & 1.02$\times 10^{-2}$  & 3.08 \\
$T_{1,\rm t}^{-1}$ (s$^{-1}$)   & 0.15                  & 32.8 \\  
$T_{1,\rm jn}^{-1}$ (s$^{-1}$)  & 15.4                  & 1.77$\times 10^4$ \\  
$T_{2,\rm jn}^{*-1}$ (s$^{-1}$) & 3.64$\times 10^{-2}$  & 2.35$\times 10^3$ \\    
\midrule[0.5pt]\bottomrule[1pt]
\end{tabular}
}
\caption{Precession frequency, dipole and gate coupling matrix elements, inverse relaxation and coherence times at the S and V points of Fig. 4, main text.}
\label{TableT}
\end{table}

\providecommand{\latin}[1]{#1}
\providecommand*\mcitethebibliography{\thebibliography}
\csname @ifundefined\endcsname{endmcitethebibliography}
  {\let\endmcitethebibliography\endthebibliography}{}


\begin{thebibliography}{36}
\bibitem [{not()}]{noteDissociate}%
  \BibitemOpen
  \href@noop {} {}\bibinfo {note} {In essence, we dissociate here the action of
  the RF field and the action of SO coupling with respect to a conventional
  EDSR set-up, where SO coupling would act on the spin during the RF
  excitation.}\BibitemShut {Stop}%
\end{thebibliography}
\end{document}